\documentclass[%
 reprint,
 amsmath,amssymb,
 aps,
]{revtex4-2}

\usepackage{graphicx}
\usepackage{dcolumn}
\usepackage{bm}
\usepackage{chemformula}

\begin{document}

\preprint{APS/123-QED}

\title{Quantifying the Spin-Orbital Entanglement in $5d^1$ Quantum Materials}

\author{V. Garc\'ia-Rojas}
 \affiliation{Escuela de Qu\'imica, Universidad Industrial de Santander, Cra. 27--9, 680002 Bucaramanga, Colombia}
\author{J. F. P\'erez-Torres}%
 \email{jfperezt@uis.edu.co}
 \affiliation{Escuela de Qu\'imica, Universidad Industrial de Santander, Cra. 27--9, 680002 Bucaramanga, Colombia}

\date{\today}

\begin{abstract}
        The spin-orbital entanglement in $5d^1$ transition metal
        ions embedded in double perovskites, where anomalous effective
        magnetic dipole moments are frequently observed, is quantified
        by the spin-orbital von Neumann entropy $\Delta S_{\rm vN}^{\rm SO}$.
        The framework is grounded on the relativistic crystal field theory,
        and is illustrated through a series of quantum materials:
        \ch{$A$2TaCl6} ($A = \ch{K}, \ch{Rb}$), \ch{$A$2MgReO6}
        ($A = \ch{Ca}, \ch{Sr}, \ch{Ba}$) and \ch{Ba2NaOsO6}, all analyzed
        in their paramagnetic phases, alongside the \ch{ReF6} molecular system.
        The entropies are derived from measurements of the optical
        $d$-$d$ transitions $\Gamma_7(t_{2g})\leftarrow\Gamma_8(t_{2g})$
        and $\Gamma_8(e_g)\leftarrow\Gamma_8(t_{2g})$, and of the
        effective magnetic dipole moment $\mu_{\rm eff}$.
        It is demonstrated that, regardless of the system, the Kramers doublet
        $\Gamma_7(t_{2g})$ exhibits no spin-orbital von Neumann entropy.
        The entropies obtained for the relativistic crystal field
        states $\Gamma_8(t_{2g})$ and $\Gamma_8(e_g)$ uncover that,
        a larger effective magnetic dipole moment can be attributed to
        a grater spin-orbital entanglement, yet paradoxically not to a
        larger spin-orbit coupling constant.
\end{abstract}

\maketitle

\section{Introduction}
Quantum materials, so named for their exotic properties that emerge from the entanglement
between different degrees of freedom, have been the subject of numerous investigations in
recent years \cite{Cava2021}.
Among the most studied materials, are double perovskite oxides of heavy transition metals, which
exhibit unusual magnetic behaviors such as unconventional magnetization, interplay between ferromagnetic
and antiferromagnetic phases, multipolar domains, to name just a few \cite{Erickson2007,Xiang2007,Chen2010,Ishikawa2019,
Hirai2019,Khomskii2021,Nguyen2021,Cong2023,Pasztorova2023,Agrestini2024,Frontini2024,Soh2024,Zivkovic2024,Nicholls2025,
Muroi2025,Shibuya2025,Iwahara2025}.
These behaviours are mainly manifested in their anomalous heat capacities and their effective magnetic
dipole moments \cite{Marjerrison2016}.
For example, in perovskites with $5d^1$ transition metals, the magnetic dipole moment
varies between 0.2 and 0.8 Bohr magnetons. These values are quite different from the expected
value of 1.73 predicted by the spin-only magnetic moment. When the crystal field theory is furnished
with the spin-orbit interaction, it results in a more complete model that predicts the splitting
of the $d$ orbitals in octahedral environments into three:
the $\Gamma_8(t_{2g})$ orbitals, the $\Gamma_7(t_{2g})$ orbitals, and the $\Gamma_8(e_g)$ orbitals,
consistent with the number of optical $d$-$d$ transitions observed in these compounds:
$\Delta_1 = \Gamma_8(e_g) \leftarrow \Gamma_8(t_{2g})$ and $\Delta_2 = \Gamma_7(t_{2g}) \leftarrow \Gamma_8(t_{2g})$.
However, for the quadruplet ground state $\Gamma_8(t_{2g})$, a null magnetic dipole moment is predicted.
The anomalous magnetic dipole moment, and thus the quantum properties that emerge from it, have been
attributed to a strong spin-orbit coupling that goes beyond zeroth-order wave functions \cite{BallhausenBook,Stamokostas2018}.
In the strong crystal field approximation \cite{GriffithBook,BallhausenBook}, the optical $d$-$d$ transitions are
described by $\Delta_1 = \xi_{nd}/2 + 10Dq$ and $\Delta_2 = 3\xi_{nd}/2$, where $\xi_{nd}$
is the spin-orbit coupling constant and $Dq$ is the strength of the crystal field.
Thus, from the $d$-$d$ transitions obtained by Resonant Inelastic X-ray Scattering (RIXS) measurements
\cite{Mitrano2024,Frontini2024}, it is possible to calculate the parameters $\xi_{nd}$ and $Dq$.
The relatively high values of $\xi_{nd}$ obtained for $5d$ metals, on the order of
0.3 to 0.6 eV, confirm the strong spin-orbit coupling. Accordingly, it is widely accepted that a higher
value of the spin-orbit coupling constant will result in greater entanglement between the
orbital and spin degrees of freedom, and therefore a greater magnetic dipole moment.
However, this hypothesis has not been confirmed.
In this work, we will quantify the degree of spin-orbit entanglement for a family of
$5d^1$ hexachlorides and double perovskite oxides using the von Neumann entropy, along
with relativistic crystal field theory (RCFT) \cite{jperez2024,jperez2025}.
The von Neumann entropy, important in quantum computing and quantum information theory \cite{NakaharaBook},
has been successfully used to quantify entanglement between different degrees of freedom, e.g.
electron-phonon \cite{Roosz2022}, electron-nucleus \cite{SanzVicario2017,Blavier2022}, electron-electron
\cite{Palii2017}, orbital-orbital \cite{Materia2024,Greene-Diniz2025}, and
spin-orbital \cite{You2015,Safaiee2017,Tang2020,Gotfryd2020} in various systems.
The RCFT, which treats the spin-orbit interaction and the electrostatic crystal field on equal
footing, predicts the correct number of $d$-$d$ transitions and a non-zero magnetic dipole moment.
We will demonstrate that the entropies correlate well with the effective magnetic dipole moment;
consequently, both parameters $\xi_{nd}$ and $Dq$, and the relativistic ratio $p/q$, characteristic
of relativistic effects beyond the spin-orbit interaction, will play an important role in the magnetic
properties of these compounds.
The paper is organized as follows: In Sec. \ref{secRCFT} the RCTF is briefly described.
Then, in Sec. \ref{secEntropy} the density matrix formalism is introduced in order to define the spin-orbital
von Neumann entropy for the $nd^1(O_h)$ relativistic crystal field states. In Sec. \ref{QMaterials}
The developed model is benchmarked against $5d^1(O_h)$ transition metal hexachlorides and double perovskite
oxides that exhibit quantum properties arising from strong spin-orbit coupling.
Finally, some conclusions and perspectives are outlined in Sec. \ref{secConclusions}.

\section{Relativistic Crystal Field Theory}\label{secRCFT}
\begin{figure}[h!]
        \includegraphics[width=0.45\textwidth]{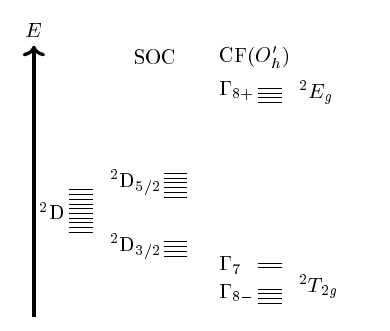}
        \caption{Energy splitting of the $\rm ^2D$ atomic state due to the spin-orbit coupling (SOC) and to the octahedral
        crystal field (CF) interactions.
        When the SOC is negligible the $\Gamma_{8-}$ and $\Gamma_7$ relativistic states, often referred to as $J_{\rm eff}=3/2$
	and $J_{\rm eff}=1/2$, respectively, resemble the $^2T_{2g}$ non-relativistic crystal field state, see Ref. \cite{Stamokostas2018}.}
        \label{fig:levels}
\end{figure}

In crystal field theory, the motion of a d-electron is represented by the Hamiltonian \cite{GriffithBook,BallhausenBook,SuganoBook}
\begin{equation}
        \hat{\cal H} = \hat{\cal H}_{\rm atom} + \hat{\cal H}_{\rm crystal-field}
\end{equation}
where $\hat{\cal H}_{\rm crystal-field}$ is the electrostatic crystal field potential \cite{GriffithBook,BallhausenBook,SuganoBook}.
In contrast to the standard crystal field theory, in its relativistic version $\hat{\cal H}_{\rm atom}$ takes the Dirac form \cite{jperez2024}:
\begin{equation}
        \hat{\cal H}_{\rm atom} = c\hat{\boldsymbol\alpha}\cdot\hat{\bf p} + \hat{\beta}mc^2 + V(r) 
\end{equation}
where $\hat{\boldsymbol \alpha}=(\alpha_x,\alpha_y,\alpha_z)$ and $\hat{\beta}$ represent the Dirac matrices,
$c$ is the velocity of the light in the vacuum, ${\bf p}$ and $m$ are the momentum and the mass of the electron, respectively,
and $V(r)$ is the potential that contains the electron-nuclei and electron-inner electrons Coulomb interactions.
For $d$-electrons, the eigenstates of $\hat{\cal H}_{\rm atom}$ correspond to four-fold $nd_{3/2}$ and six-fold $nd_{5/2}$ energy
degenerated spinors with magnetic quantum numbers $-3/2\le m_j \le +3/2$ and $-5/2 \le m_j \le +5/2$, respectively \cite{GreinerBook}.
When the atom is placed in a solid, the degeneracy is partially lifted because of the interaction with the surroundings, represented by
$\hat{\cal H}_{\rm crystal-field}$.
The interaction is mainly due to nearest neighbor atoms, the so-called ligands. In the case of the most common six-coordinated
octahedral symmetry, the six spinors $nd_{5/2}$ split into two sets of spinors which span the irreducible representations
$\Gamma_7$ and $\Gamma_8$ of the double point group $O_h^\prime$ \cite{Bethe1929}. Since the $nd_{3/2}$ set also span the
irreducible representation $\Gamma_8$, the two sets of spinors belonging to $\Gamma_8$ mixes up \cite{Stamokostas2018}.
Taking advace of the block diagonal matrix representation of the total Hamiltonian in the basis of symmetrized atomic Dirac spinors
$\{nd_{3/2}\cup nd_{5/2}\}$, the crystal field eigenstates are found to be \cite{jperez2025},
\begin{eqnarray}
	|\Gamma_7^a\rangle &=& \sqrt{\tfrac{1}{6}}|n{-3}{-\tfrac{5}{2}}\rangle -\sqrt{\tfrac{5}{6}}|n{-3}{+\tfrac{3}{2}}\rangle \label{eq:G7a} \\
	|\Gamma_7^b\rangle &=& \sqrt{\tfrac{1}{6}}|n{-3}{+\tfrac{5}{2}}\rangle - \sqrt{\tfrac{5}{6}}|n{-3}{-\tfrac{3}{2}}\rangle \label{eq:G7b} \\
	|\Gamma_{8\pm}^a\rangle &=& x_\pm\left(\sqrt{\tfrac{5}{6}}|n{-3}{-\tfrac{5}{2}}\rangle
                                        +\sqrt{\tfrac{1}{6}}|n{-3}{+\tfrac{3}{2}}\rangle\right) \nonumber \\
				  &&    +y_\pm|n{+2}{+\tfrac{3}{2}}\rangle \\
	|\Gamma_{8\pm}^b\rangle &=& x_\pm\left(\sqrt{\tfrac{5}{6}}|n{-3}{+\tfrac{5}{2}}\rangle
                                        +\sqrt{\tfrac{1}{6}}|n{-3}{-\tfrac{3}{2}}\rangle\right) \nonumber \\
				  &&    -y_\pm|n{+2}{-\tfrac{3}{2}}\rangle \\
	|\Gamma_{8\pm}^c\rangle &=& x_\pm|n{-3}{-\tfrac{1}{2}}\rangle + y_\pm|n{+2}{-\tfrac{1}{2}}\rangle \\
	|\Gamma_{8\pm}^d\rangle &=& x_\pm|n{-3}{+\tfrac{1}{2}}\rangle - y_\pm|n{+2}{+\tfrac{1}{2}}\rangle
\end{eqnarray}
with expansion coefficients
\begin{eqnarray}
        && x_\pm = \frac{\delta\pm\sqrt{\delta^2+1}}{\sqrt{\left(\delta\pm\sqrt{\delta^2+1}\right)^2+1}} \\
        && y_\pm = \frac{1}{\sqrt{\left(\delta\pm\sqrt{\delta^2+1}\right)^2+1}} \\
        && \delta = \frac{1}{2\sqrt{6}(p/q)}\left(\frac{5\xi_{nd}}{4Dq}+1\right) \label{eq:delta}
\end{eqnarray}
effective magnetic dipole moment
\begin{eqnarray}
	&& \mu_{\rm eff} = \sqrt{\frac{3}{4}\sum_{m_j=-3/2}^{3/2}(W_{m_j}^{\rm I})^2} \label{eq:mueff} \\
	&& W_{\pm1/2}^{\rm I} = \pm\frac{1}{2}\left(\frac{6}{5}x_-^2 + \frac{4}{5}y_-^2 + \frac{4\sqrt{6}}{5}x_-y_-\right)\mu_{\rm B} \\
        && W_{\pm3/2}^{\rm I} = \pm\frac{3}{2}\left(\frac{4}{5}y_-^2 - \frac{22}{15}x_-^2 - \frac{4\sqrt{6}}{45}x_-y_-\right)\mu_{\rm B}
\end{eqnarray}
and energy levels
\begin{eqnarray}
        && E(\Gamma_7) = \xi_{nd} - 4Dq \label{eq:E7} \\
        && E(\Gamma_{8\pm}) = -\frac{\xi_{nd}}{4} + \left(1 \pm \sqrt{\left(\frac{5\xi_{nd}}{4Dq}+1\right)^2
        + 24(p/q)^2}\right)Dq \nonumber \\ \label{eq:E8}
\end{eqnarray}
Here $\pm$ stands for the lower $(-)$ and upper $(+)$ energy levels $\Gamma_8(t_{2g})$ and $\Gamma_8(e_g)$, respectively,
see Figure \ref{fig:levels}.
The ket notation $|n\kappa m_j\rangle$ has been used to represent the atomic spinors $d_{3/2}$ ($\kappa=+2$) and $d_{5/2}$ ($\kappa=-3$).
The spin-orbit coupling constant $\xi_{nd}$, the crystal field strength $Dq$, and the relativistic ratio $p/q$,
defined as \cite{jperez2024}
\begin{eqnarray}
        && \xi_{nd} = \frac{2}{5}\left(E(nd_{5/2}) - E(nd_{3/2})\right) \\
        && Dq = \alpha\hbar c \frac{Z}{6a} \int_0^\infty |F_{nd_{5/2}}(r)|^2 r^4 {\rm d}r \\
        && p/q = \frac{\int_0^\infty F_{nd_{5/2}}(r)F_{nd_{3/2}}(r)r^4{\rm d}r}{\int_0^\infty |F_{nd_{5/2}}(r)|^2 r^4 {\rm d}r}
        \label{eq:pq}
\end{eqnarray}
can be obtained from the optical spectrum ($d$-$d$ transitions) supplemented by magnetic susceptibility measurements \cite{jperez2025}
or ab-initio relativistic quantum chemistry calculations \cite{jperez2024}. The radial functions $F_{nd_j}(r)$
correspond to the large component of the atomic spinors (see Appendix), $\alpha$ is the fine structure constant,
$\hbar$ is the reduced planck constant, $\mu_{\rm B}$ is the Bohr magneton, $Z$ is the effective charge of the ligands,
and $a$ is the ligand-metal distance. We note that if we set $p/q=1$, the present formalism becomes equivalent to the full
$t_{2g}$-$e_g$ one-electron model developed in Ref. \cite{Stamokostas2018}.
Using the Taylor expansion for $\sqrt{1+x}$ in Eq. \eqref{eq:E8} and assuming $10Dq\gg\xi_{nd}$ and $p/q=1$,
it can be shown that
\begin{eqnarray}
        && \Delta_1 = E(\Gamma_{8+})-E(\Gamma_{8-}) \approx \tfrac{1}{2}\xi_{nd} + 10Dq \\
        && \Delta_2 = E(\Gamma_7)-E(\Gamma_{8-}) \approx \tfrac{3}{2}\xi_{nd}
\end{eqnarray}
which is known as the strong field approximation \cite{Moffitt1959,Rotger1999}.
Moreover, if $\xi_{nd}\to0$, then both radial functions $F_{nd_{5/2}}(r)/r$ and $F_{nd_{3/2}}(r)/r$ are equal
to the radial function $R_{nd}(r)$ of the Schr\"odinger orbitals $\psi_{n\ell m}({\bf r})=R_{n\ell}(r)Y_{\ell m}(\theta\phi)$,
and therefore $p=q$ is obtained, see Eq. \eqref{eq:pq}.
Then, it follows that $E(\Gamma_{8-}) = E(\Gamma_{7}) = -4Dq$ and $E(\Gamma_{8+}) = +6Dq$, matching up with the energy
levels $E(^2T_{2g})$ and $E(^2E_g)$ of the non-relativistic crystal field theory.
Note that the pair of states $|\Gamma_i^a\rangle$ and $|\Gamma_i^b\rangle$, and $|\Gamma_i^c\rangle$ and $|\Gamma_i^d\rangle$, comprise
Kramers doublets, i.e. $|\Gamma_i^b\rangle = -\hat{\cal T}|\Gamma_i^a\rangle$ and $|\Gamma_i^d\rangle = -\hat{\cal T}|\Gamma_i^c\rangle$,
where $\hat{\cal T}:t\to-t\equiv \hat{\cal T}|jm_j\rangle=(-1)^{j-m_j}|j-m_j\rangle$ is the time-reversal symmetry
operator \cite{GreinerBook}.
The theoretical framework described above has been used to analyze the optical and magnetic properties of $4d^1$ and $5d^1$
transition metal ions in cubic and tetragonal ligand fields \cite{jperez2024,jperez2025}.

\section{Density Matrices and von Neumann Entropy}\label{secEntropy}
\begin{figure}
\includegraphics[width=0.45\textwidth]{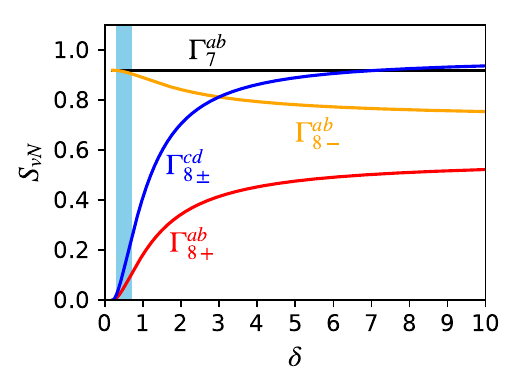}
\caption{von Neumann entropy $S_{\rm vN}$ of the $\Gamma_7$ and $\Gamma_8$ states of $nd^1$ ions in octahedral
        crystal fields as a function of $\delta$.
        The shadow rectangle indicates the typical values of $\delta$ for $5d^1(O_h)$ systems ($0.3\le\delta\le0.7$).}
        \label{fig:SvN}
\end{figure}

Suppose we have a system described by the pure quantum state $|\Gamma_i\rangle$, then the density matrix operator associated to
the state is given by \cite{NakaharaBook}
\begin{eqnarray}
        \hat{\rho}_{\Gamma_i} = |\Gamma_i\rangle\langle\Gamma_i|
\end{eqnarray}
If the quantum state $|\Gamma_i\rangle$ contains orbital angular momentum $\ell$ and spin angular momentum $s$ degrees of freedom,
the partial traces for $\ell=2$ and $s=1/2$ Hilbert subspaces,
\begin{eqnarray}
        && {\rm Tr}_\ell\left(\hat{\rho}_{\Gamma_i}\right)
                                          = \sum_{m_\ell=-2}^2\langle2m_\ell|\hat{\rho}_{\Gamma_i}|2m_\ell\rangle
                                          = \hat{\rho}_{\Gamma_i}^{(s)}\\
        && {\rm Tr}_s\left(\hat{\rho}_{\Gamma_i}\right)
                                          = \sum_{m_s=-1/2}^{1/2}\langle \tfrac{1}{2}m_s|\hat{\rho}_{\Gamma_i}|\tfrac{1}{2}m_s\rangle
                                          = \hat{\rho}_{\Gamma_i}^{(\ell)}
\end{eqnarray}
define the reduced density matrix operators $\hat{\rho}_{\Gamma_i}^{(s)}$ and $\hat{\rho}_{\Gamma_i}^{(\ell)}$ associated
to the spin and orbital angular momentum subspaces, respectively. Both reduced density matrix operators admit matrix
representations in the eigenbasis $|\tfrac{1}{2}m_s\rangle$ and $|2m_\ell\rangle$, that is
$\rho^{(s)}_{m_s^\prime m_s} = \langle \tfrac{1}{2}m_s^\prime|\hat{\rho}^{(s)}_{\Gamma_i}|\tfrac{1}{2}m_s\rangle$ and
$\rho^{(\ell)}_{m_\ell^\prime m_\ell} = \langle 2m_\ell^\prime|\hat{\rho}^{(\ell)}_{\Gamma_i}|2m_\ell\rangle$,
and share the same spectrum \cite{SanzVicario2017}.
The spectrum of a reduced density matrix of dimension $N$ is the set of eigenvalues $\{\lambda_i\}$ that satisfy
$0\le\lambda_i\le1$ and $\sum_{i=1}^N\lambda_i=1$.
If the spectrum is such that $\lambda_i=\delta_{ij}$, then the quantum state is said to be separable; otherwise, the quantum
state is said to be entangled. Finally, the von Neumann entropy, defined as
\begin{eqnarray}
        S_{\rm vN} = -{\rm Tr}(\hat{\rho}\log_2\hat{\rho}) = -\sum_i\lambda_i\log_2(\lambda_i)
\end{eqnarray}
quantifies the strength of entanglement between the two halfspaces.
It turns out that, for our particular case of $nd^1(O_h)$ systems, the matrix representation of the reduced density
matrices $\hat{\rho}_{\Gamma_i}^{(s)}$ of the spin halfspace is diagonal in the basis of Pauli eigenvectors
$\{|\alpha\rangle=|\tfrac{1}{2}{+\tfrac{1}{2}}\rangle, |\beta\rangle=|\tfrac{1}{2}{-\tfrac{1}{2}}\rangle\}$.
Therefore, the eigenvalues required to obtain the von Neumann entropies of the quantum states $\Gamma_i$ turn out to
be:
\begin{eqnarray}
	\lambda_1(\Gamma_7^a) &=& \langle\alpha|\hat{\rho}_{\Gamma_{7a}}^{(s)}|\alpha\rangle = \frac{2}{3} \\
	\lambda_2(\Gamma_7^a) &=& \langle\beta|\hat{\rho}_{\Gamma_{7a}}^{(s)}|\beta\rangle = \frac{1}{3} \\
	\lambda_1(\Gamma_{8\pm}^a) &=& \langle\alpha|\hat{\rho}_{\Gamma_{8\pm a}}^{(s)}|\alpha\rangle \nonumber \\
				   &=& \frac{2}{15}x_\pm^2 + \frac{1}{5}y_\pm^2 - \frac{2\sqrt{6}}{15}x_\pm y_\pm \\
	\lambda_2(\Gamma_{8\pm}^a) &=& \langle\beta|\hat{\rho}_{\Gamma_{8\pm a}}^{(s)}|\beta\rangle \nonumber \\
				   &=& \frac{13}{15}x_\pm^2 + \frac{4}{5}y_\pm^2 + \frac{2\sqrt{6}}{15}x_\pm y_\pm \\
	\lambda_1(\Gamma_{8\pm}^c) &=& \langle\alpha|\hat{\rho}_{\Gamma_{8\pm c}}^{(s)}|\alpha\rangle \nonumber \\
				   &=& \frac{2}{5}x_\pm^2 + \frac{3}{5}y_\pm^2 - \frac{2\sqrt{6}}{5}x_\pm y_\pm \\
	\lambda_2(\Gamma_{8\pm}^c) &=& \langle\beta|\hat{\rho}_{\Gamma_{8\pm c}}^{(s)}|\beta\rangle \nonumber \\
				   &=& \frac{3}{5}x_\pm^2 + \frac{2}{5}y_\pm^2 + \frac{2\sqrt{6}}{5}x_\pm y_\pm
\end{eqnarray}
The eigenvalues for the remaining states can be obtained using the time-reversal symmetry operator, e.g.
$\lambda_{1}(\Gamma_7^b) = \langle\alpha|\hat{\cal T}^{-1}\hat{\cal T}|\rho^{\rm (s)}_{\Gamma_7^b}|
\hat{\cal T}\hat{\cal T}^{-1}|\alpha\rangle$.
Notice that $\lambda_1(\Gamma_i)+\lambda_2(\Gamma_i)=1$. As expected from Eqs. \ref{eq:G7a} and \ref{eq:G7b},
the von Neumann entropy for the crystal field states $\Gamma_7^a$ and $\Gamma_7^b$ is constant:
$S_{\rm vN}(\Gamma_7^{ab})=-(1/3)\log_2(1/3)-(2/3)\log_2(2/3)=0.918$, that is, it is the same for any compound $nd^1(O_h)$.
Since the parameter $\delta$ contains all the experimental quantities $Dq$, $\xi_{nd}$ and $p/q$,
it is convenient to represent the entropies $S_{\rm vN}(\Gamma_i)$ as a function of $\delta$.
Figure \ref{fig:SvN} shows the von Neumann entropy for the Kramers doublets $(\Gamma_{8\pm}^a,\Gamma_{8\pm}^b)$,
$(\Gamma_{8\pm}^c,\Gamma_{8\pm}^d)$, and $(\Gamma_7^a,\Gamma_7^b)$.
Surprisingly, $S_{\rm vN}(\Gamma_{8-}^{cd}) = S_{\rm vN}(\Gamma_{8+}^{cd})$, i.e. two Kramers doublets from different energy levels
share the same degree of spin-orbital entanglement. In general, $S_{\rm vN}(\Gamma_{8-}^{ab})$ barely changes with $\delta$, while
$S_{\rm vN}(\Gamma_{8\pm}^{cd})$ and $S_{\rm vN}(\Gamma_{8+}^{ab})$ change abruptly.
In the limit of zero spin-orbit interaction, i.e. $\xi_{nd}\to0$, we obtain $p=q$, $\delta = 1/2\sqrt{6}$
(see Eqs. \eqref{eq:pq} and \eqref{eq:delta}), $S_{\rm vN}(\Gamma_7^{ab})=S_{\rm vN}(\Gamma_{8-}^{ab})=0.918$
and $S_{\rm vN}(\Gamma_{8\pm}^{cd})=S_{\rm vN}(\Gamma_{8+}^{ab})=0$. This can be understood as follows: for zero spin-orbit
interaction, the relativistic crystal field states $|\Gamma_i\rangle$ resemble, up to a global phase, the Ballhausen's
crystal field spin-orbitals \cite{BallhausenBook}, see Table \ref{tab:RCFT}. It immediately follows that only the states
$\Gamma_{8-}^{ab}$ and $\Gamma_7^{ab}$ are entangled (non-separable) states.
Indeed, von Neumann entropy quantifies our ignorance about the quantum state of a system. For instance, if the system is
described by the state $\Gamma_{8^c}(t_{2g})$, we know with certainty that the electron is represented
by the crystal field orbital $t_{2g}^-$ with spin state $\alpha$, and therefore $S_{\rm vN}(\Gamma_{8^c}(t_{2g}))=0$.
Additionally, the eigenvalues of the density matrices of the entangled states are easily obtained:
$\lambda_1(\Gamma_{8^a})=\lambda_2(\Gamma_{7^a})=1/3$ and $\lambda_2(\Gamma_{8^a})=\lambda_1(\Gamma_{7^a}) = 2/3$,
in agreement with our early results.
Obviously, such entanglement can not be attributed to the spin-orbit interaction but to the symmetry point group.
Therefore, in order to quantify the spin-orbital entanglement, we define the spin-orbital von Neumann entropy for
the state $\Gamma_i$ as
\begin{equation}
\Delta S_{\rm vN}^{\rm SO}(\Gamma_i)=S_{\rm vN}(\Gamma_i)-S_{\rm vN}^{(\xi_{nd}=0)}(\Gamma_i)
\end{equation}
Thus, the spin-orbit interaction does not produce any entropy in the Kramers doublet $\Gamma_7$.
Only the Kramers doublets $\Gamma_8$ are influenced by the spin-orbit interaction, with negative spin-orbital entropy
for the Kramers doublet $\Gamma_{8-}^{ab}$ as the spin-orbit interaction is turned on.
\begin{table}[h!]
        \caption{Correlation between Ballhausen's spin-orbitals \cite{BallhausenBook} and relativistic crystal field states
        (RCFS) for $\xi_{nd}\to0$. For a detailed description of the spin-orbitals see page 118 of
        Ballhausen's Book \cite{BallhausenBook}.}\label{tab:RCFT}
        \begin{tabular}{ll}\hline\hline
		Ballhausen's spin-orbitals & RCFS\\ \hline
		$\Gamma_{8^a}(t_{2g}) = \sqrt{1/3}(-\sqrt{2}t_{2g}^0\beta + t_{2g}^+\alpha)$ 	 &  $-\Gamma_{8-}^a$   	\\
		$\Gamma_{8^b}(t_{2g}) = \sqrt{1/3}(-\sqrt{2}t_{2g}^0\alpha - t_{2g}^-\beta)$ 	 &  $\Gamma_{8-}^b$      \\
		$\Gamma_{8^c}(t_{2g}) = t_{2g}^-\alpha$ 					 &  $-\Gamma_{8-}^c$     \\
		$\Gamma_{8^d}(t_{2g}) = -t_{2g}^+\beta$ 					 &  $\Gamma_{8-}^d$      \\
                $\Gamma_{7^a}(t_{2g}) = \sqrt{1/3}(t_{2g}^0\beta + \sqrt{2}t_{2g}^+\alpha)$ 	 &  $-\Gamma_7^a$      \\
		$ \Gamma_{7^b}(t_{2g}) =  \sqrt{1/3}(-t_{2g}^0\alpha + \sqrt{2}t_{2g}^-\beta)$ 	 &  $\Gamma_7^b$       \\
		$ \Gamma_{8^a}(e_g) =  e_g^b\beta$ 						 &  $\Gamma_{8+}^a$      \\
		$ \Gamma_{8^b}(e_g) =  -e_g^b\alpha$ 						 &  $-\Gamma_{8+}^b$     \\
		$ \Gamma_{8^c}(e_g) =  e_g^a\beta$ 						 &  $\Gamma_{8+}^c$      \\
		$ \Gamma_{8^d}(e_g) =  -e_g^a\alpha$ 						 &  $-\Gamma_{8+}^d$     \\
                \hline\hline
        \end{tabular}
\end{table}
The asymptotic values of $S_{\rm vN}$ for large $\delta$, i.e. $Dq\to0$, are easily obtained from the limits
\begin{eqnarray}
        && \lim_{Dq\to0}x_- = \lim_{Dq\to0}y_+=0 \\
        && \lim_{Dq\to0}x_+=\lim_{Dq\to0}y_-=1
\end{eqnarray}
which give rise to the asymptotic entropies
\begin{eqnarray}
	S_{\rm vN}(\Gamma_{8-}^{ab})   &=&-\tfrac{1}{5}\log_2(\tfrac{1}{5})-\tfrac{4}{5}\log_2(\tfrac{4}{5}) \nonumber \\
					&=& 0.722 \\
	S_{\rm vN}(\Gamma_{8+}^{ab})   &=& -\tfrac{2}{15}\log_2(\tfrac{2}{15})-\tfrac{13}{15}\log_2(\tfrac{13}{15}) \nonumber \\
					&=& 0.567 \\
	S_{\rm vN}(\Gamma_{8\pm}^{cd}) &=& -\tfrac{2}{5}\log_2(\tfrac{2}{5})-\tfrac{3}{5}\log_2(\tfrac{3}{5}) \nonumber \\
					&=& 0.971
\end{eqnarray}
Remarkably, none of the states reaches the maximum entanglement $\lambda_1^{\rm max} = \lambda_2^{\rm max}=1/2$ and
$S_{\rm vN}^{\rm max} = 1$.
From the monotonic behavior of $\Delta S_{\rm vN}^{\rm SO}$ with $\delta$, follows that the spin-orbit entanglement
increases with the spin-orbit coupling constant $\xi_{nd}$, as expected, but decreases with the crystal field
strength $Dq$ and with the ratio $p/q$.

\section{Spin-orbital von Neumann Entropy in Quantum Materials}\label{QMaterials}
\begin{figure}
\includegraphics[width=0.45\textwidth]{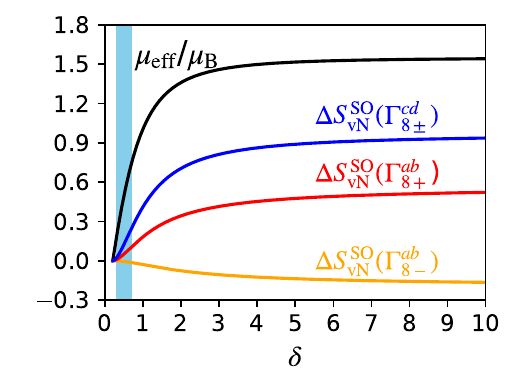}
        \caption{Effective magnetic dipole moment $\mu_{\rm eff}$ and spin-orbital von Neumann entropies $\Delta S_{\rm vN}^{\rm SO}$
        for $nd^1$ ions in octahedral crystal fields as a function of the parameter $\delta$.
        The shadow rectangle indicates the typical values of $\delta$ for $5d^1(O_h)$ systems ($0.3\le\delta\le0.7$).}
        \label{fig:muSvN}
\end{figure}

In this section we study the spin-orbital entanglement in a set of quantum materials through the spin-orbital
von Neumann entropy.
The family of hexachlorides \ch{$A$2TaCl6} ($A = \ch{K}, \ch{Rb}$) and double perovskite oxides \ch{$A$2MgReO6}
($A = \ch{Ca}, \ch{Sr}, \ch{Ba}$) and \ch{Ba2NaOsO6}, are employed as benchmarks.
The \ch{ReF6} molecular system is analyzed as well.
We should point out that the local symmetry of \ch{$A$2MgReO6} and \ch{Ba2NaOsO6} is $D_{4h}$ and not $O_h$,
due to dynamic Jahn-Teller effect \cite{Agrestini2024,Frontini2024}.
However, the spliting of the ground state $\Gamma_8(t_{2g})$ is one order of magnitude smaller than the spin-orbit
coupling, being $\Delta_{\rm JT}\approx0.030~{\rm eV}$ for \ch{$A$2MgReO6} and similar for \ch{Ba2NaOsO6}.
On the other hand, the RIXS spectra at room temperature reveal a four-fold degenerated $\Gamma_8(t_{2g})$ ground state
for the family \ch{$A$2TaCl6} ($A = \ch{K}, \ch{Rb}, \ch{Cs}$) \cite{Ishikawa2019}.
Therefore, assuming octahedral symmetry in all cases is safe.
Table \ref{tab:data} gathers the reported experimental optical $d$-$d$ transition energies and effective magnetic dipole moments
used in this work. The calculated spin-orbit coupling constants $\xi_{5d}$, crystal field strengths $Dq$,
and ratios $p/q$, were obtained according to the protocol described in ref \cite{jperez2025}, and are reported in Table \ref{tab:SvN}
together with the spin-orbital von Neumann entropies $\Delta S_{\rm vN}^{\rm SO}$.
Because of the lack of reported magnetic susceptibility measurements for \ch{Ca2MgReO6} and \ch{Sr2MgReO6}, ab-initio relativistic quantum
chemistry calculations were carried out for atomic \ch{Re^6+} in order to obtain the ratio $p/q=0.963$ as reported in ref \cite{jperez2024}.
\begin{table}
        \caption{Reported $d$-$d$ optical transitions
        $\Delta_1$: $\Gamma_8(e_g) \leftarrow \Gamma_8(t_{2g})$,
        $\Delta_2$: $\Gamma_7(t_{2g}) \leftarrow \Gamma_8(t_{2g})$,
        and effective magnetic dipole moments
        $\mu_{\rm eff}$ employed in this work.}\label{tab:data}
        \begin{tabular}{lccc}\hline\hline
                               & $\Delta_1/{\rm eV}$             & $\Delta_2/{\rm eV}$            & $\mu_{\rm eff}/\mu_{\rm B}$ \\ \hline
                \ch{K2TaCl6}   & 3.20 (ref \cite{Ishikawa2019})  & 0.40 (ref \cite{Ishikawa2019}) & 0.30 (ref \cite{Ishikawa2019}) \\
                \ch{Rb2TaCl6}  & 3.20 (ref \cite{Ishikawa2019})  & 0.40 (ref \cite{Ishikawa2019}) & 0.27 (ref \cite{Ishikawa2019}) \\
                \ch{ReF6}      & 3.65 (ref \cite{Rotger1999})    & 0.62 (ref \cite{Rotger1999})   & 0.25 (ref \cite{Selig1962}) \\
                \ch{Ca2MgReO6} & 4.95 (ref \cite{Frontini2024})  & 0.545(ref \cite{Frontini2024}) &     --      \\
                \ch{Sr2MgReO6} & 4.91 (ref \cite{Frontini2024})  & 0.535(ref \cite{Frontini2024}) &     --      \\
                \ch{Ba2MgReO6} & 4.56 (ref \cite{Frontini2024})  & 0.52 (ref \cite{Frontini2024}) & 0.68 (ref \cite{Hirai2019}) \\
                \ch{Ba2NaOsO6} & 4.86 (ref \cite{Agrestini2024}) & 0.51 (ref \cite{Agrestini2024})& 0.677 (ref \cite{Stitzer2002}) \\
                \hline\hline
        \end{tabular}
\end{table}
\begin{table}[h!]
        \caption{Spin-orbit coupling constant $\xi_{5d}$, crystal field stength $Dq$, relativistic ratio $p/q$,
        and spin-orbital von Neumann entropies $\Delta S_{\rm vN}^{\rm SO}$ of crystal field states of $5d^1$
        transition metal ions in octahedral environments.
        The values are obtained using the effective magnetic dipole moments and energies presented in the Table \ref{tab:data},
        see Appendix.
        *Values obtained from ab-initio relativistic quantum chemistry calculations at the Dirac-Hartree-Fock level \cite{jperez2024}.}
        \label{tab:SvN}
        \begin{tabular}{lcccccc}\hline\hline
		& $\xi_{5d}/{\rm eV}$ & $Dq/{\rm eV}$ & $p/q$ & \multicolumn{3}{c}{$\Delta S_{\rm vN}^{\rm SO}$} \\
		& & & & $\Gamma_{8-}^{ab}$ & $\Gamma_{8+}^{ab}$ & $\Gamma_{8\pm}^{cd}$ \\ \hline
                \ch{K2TaCl6}    & 0.211 & 0.293 & 1.046  & $-0.0020$ & 0.0205 & 0.0522 \\
                \ch{Rb2TaCl6}   & 0.195 & 0.289 & 1.066  & $-0.0016$ & 0.0170 & 0.0435 \\
                \ch{ReF6}       & 0.232 & 0.299 & 1.180  & $-0.0013$ & 0.0148 & 0.0379 \\
                \ch{Ca2MgReO6}  & 0.386 & 0.483 & 0.963* & $-0.0034$ & 0.0321 & 0.0804 \\
                \ch{Sr2MgReO6}  & 0.380 & 0.479 & 0.963* & $-0.0033$ & 0.0316 & 0.0794 \\
                \ch{Ba2MgReO6}  & 0.584 & 0.498 & 0.787  & $-0.0115$ & 0.0882 & 0.2112 \\
                \ch{Ba2NaOsO6}  & 0.614 & 0.538 & 0.778  & $-0.0114$ & 0.0874 & 0.2094 \\
                \hline\hline
        \end{tabular}
\end{table}
In general, the spin-orbit coupling constants obtained here follow the expected trend with the atomic number $Z$
for isoelectronic ions, i.e. $\xi_{5d}(\ch{Ta^4+}) < \xi_{5d}(\ch{Re^6+}) < \xi_{5d}(\ch{Os^7+})$.
To contrast the results, we compare the magnetic moments of \ch{K2TaCl6} and \ch{Ba2NaOsO6}, as reported in Table 2,
with the spin-orbit coupling values calculated in Table 3. For this pair, it seems to be a clear correlation indicating that
greater spin-orbit coupling corresponds to a higher magnetic moment. However, when we examine the pairs \ch{K2TaCl6}
and \ch{ReF6}, a different trend emerges; although the magnetic moment of \ch{K2TaCl6} is greater than that of \ch{ReF6},
the calculated spin-orbit coupling for \ch{ReF6} is actually higher than that for \ch{K2TaCl6}.
However, within the families \ch{$A$2TaCl6} and \ch{$A$2MgReO6}, it is found the spin-orbit interaction takes precedence
over the crystal field strength, since the spin-orbital von Neumann entropy follows the expected behavior with $\xi_{5d}$
but not with $Dq$. Thus, in general, none of the crystal field parameters correlates with the entropy and therefore with the
degree of spin-orbital entanglement.
Interestingly, despite the different spin-orbit coupling constants for \ch{Ba2MgReO6} and \ch{Ba2NaOsO6}, the von Neumann
entropies reveal comparable, if not equal, entanglement for both perovskites.
This is so because the increase in $\xi_{5d}$ is offset by the increase in $Dq$ at almost equal $p/q$, equilibrating the $\delta$
values for both perovskites.
This result confirms our expectation that, to quantify the spin-orbital entanglement, all the relativistic
crystal field parameters must be taken into account. This is achieved through the spin-orbital von Neumann entropy.
Moreover, a notable correlation between the entropies and the effective magnetic dipole moment is observed from
Tables \ref{tab:data} and \ref{tab:SvN}. This can be better appreciated from Figure \ref{fig:muSvN}, where the
effective magnetic dipole moment and the spin-orbital von Neumann entropies are plotted against the parameter $\delta$.
Note that $\delta$ contains all the relativistic crystal field parameters, see eqn \eqref{eq:delta}.
Thus, the relativistic ratio and the crystal field strength also play a role in the entanglement between the orbital and spin
degrees of freedom and can not be ignored.
This findings confirm that the anomalous effective magnetic dipole moment, attributed to the strong spin-orbital entanglement,
is better characterized by the spin-orbital von Neumann entropy rather than the spin-orbit coupling constant.
We should mention that the vanishing of the effective magnetic dipole moment when $\delta\to1/2\sqrt{6}$ must be associated to
$\xi_{nd}/Dq\to0$ instead of $\xi_{nd}\to0$. This is because, as the spin-orbit coupling constant decreases,
on the order of $k_{\rm B}T\approx0.0259~{\rm eV}$, the van Vleck magnetic susceptibility \cite{vanVleck1978} caused by the
virtual transitions between the states $\Gamma_{8-}$ and $\Gamma_7$ becomes significant, for which Eq. \eqref{eq:mueff} is not
longer valid.
Finally, because of the entanglement between the spin, orbital, and lattice degrees of freedom
\cite{Frontini2024,Iwahara2025,Muroi2025}, a small effect of the temperature on the entropies should be expected.

\section{Conclusions and outlook}\label{secConclusions}
In conclusion, it has established a framework that allows us to quantify the spin-orbital entanglement in the quantum
states of $nd^1(O_h)$ systems from optical and magnetic measurements.
The method is based on the von Neumann entropy and on the relativistic crystal field theory, yielding the spin-orbital
von Neumann (SOvN) entropy as the key quantity.
A family of $5d^1$ transition metal hexachlorides and double perovskite oxides, exhibiting properties of quantum materials,
was analyzed. The molecular system \ch{ReF6} was analyzed as well.
It was discovered that the effective magnetic dipole moment $\mu_{\rm eff}$ correlates well with the SOvN
entropy, but not necessarily with the spin-orbit coupling constant $\xi_{5d}$ or the crystal field strength $Dq$.
In general, the quadruplet ground state of these systems is characterized by two well-distinct Kramers doublets:
one doublet is weakly entangled showing negative SOvN entropy $\Delta S_{\rm vN}^{\rm SO}(\Gamma_{8-}^{ab}) \lesssim 0$,
while the other is strongly entangled with an SOvN entropy equal to that of one of the Kramers doublets of the second excited state,
i.e. $\Delta S_{\rm vN}(\Gamma_{8-}^{cd}) = \Delta S_{\rm vN}(\Gamma_{8+}^{cd})>0$.
The second Kramers doublet of the second excited state exhibits a moderate entanglement $\Delta S_{\rm vN}(\Gamma_{8+}^{ab}) \gtrsim 0$.
Curiously, the first excited state of these systems lacks SOvN entropy $\Delta S_{\rm vN}^{\rm SO}(\Gamma_7^{ab})=0$.
The phenomenological framework developed here should be considered as a bridge between experiment and theory. It would
be interesting to derive the corresponding SOvN entropy from ab initio and density functional theory calculations
\cite{Agrestini2024,Merkel2024,FioreMosca2024,Sunaga2024,Kukusta2025,Okamoto2025}, in particular employing the ab initio ligand field
theory \cite{Haverkort2012,Jung2017,Singh2017,Rao2024,Cardot2024,Nielsen2025}.
We envision that with more experimental data available, new features will emerge from the SOvN entropy.
An important prospect for the von Neumann entropy, along with the relativistic crystal field theory, lies in its possible
extensions to more complex scenarios like spin-orbital-lattice and many-electron entangled systems, including other symmetries
like tetragonal or trigonal crystal fields.

\appendix
\section{Appendix}\label{secAppendix}

\subsection{Density matrix operators of crystal field states}
\begin{widetext}
The density matrix operators for the crystal field states of $nd^1(O_h)$ ions in the basis of atomic d-spinors are
\begin{eqnarray}
        \hat{\rho}_{\Gamma_7^a} &=& \tfrac{5}{6}|n{-3}{\tfrac{3}{2}}\rangle\langle n{-3}{\tfrac{3}{2}}|
                                + \tfrac{1}{6}|n{-3}{-\tfrac{5}{2}}\rangle\langle n{-3}{-\tfrac{5}{2}}|
                                - \tfrac{\sqrt{5}}{6}|n{-3}{\tfrac{3}{2}}\rangle\langle n{-3}{-\tfrac{5}{2}}|
                                - \tfrac{\sqrt{5}}{6}|n{-3}{-\tfrac{5}{2}}\rangle\langle n{-3}{\tfrac{3}{2}}| \\
	\hat{\rho}_{\Gamma_{8\pm}^a} &=& x_\pm^2\Big{(}\tfrac{5}{6}|n{-3}{-\tfrac{5}{2}}\rangle\langle n{-3}{-\tfrac{5}{2}}|
                                        +\tfrac{1}{6}|n{-3}\tfrac{3}{2}\rangle\langle n{-3}\tfrac{3}{2}|
                                        +\sqrt{\tfrac{5}{6}}|n{-3}{-\tfrac{5}{2}}\rangle\langle n{-3}\tfrac{3}{2}|
                                        +\sqrt{\tfrac{5}{6}}|n{-3}\tfrac{3}{2}\rangle\langle n{-3}{-\tfrac{5}{2}}|\Big{)} \nonumber \\
                                        && + y_\pm^2|n2\tfrac{3}{2}\rangle\langle n2\tfrac{3}{2}|
                                           +x_\pm y_\pm \left(\sqrt{\tfrac{5}{6}}|n{-3}{-\tfrac{5}{2}}\rangle\langle
                                        n2{\tfrac{3}{2}}| + \sqrt{\tfrac{1}{6}}|n{-3}{\tfrac{3}{2}}\rangle\langle n2\tfrac{3}{2}|\right)
                                        \nonumber \\
                                        && +x_\pm y_\pm \left(\sqrt{\tfrac{5}{6}}|n2\tfrac{3}{2}\rangle\langle n{-3}{-\tfrac{5}{2}}|
                                        +\sqrt{\tfrac{1}{6}}|n2\tfrac{3}{2}\rangle\langle n{-3}\tfrac{3}{2}|\right) \\
	\hat{\rho}_{\Gamma_{8\pm}^c} &=& x_\pm^2|n{-3}{-\tfrac{1}{2}}\rangle\langle n{-3}{-\tfrac{1}{2}}|
                                      + y_\pm^2|n2{-\tfrac{1}{2}}\rangle\langle n2{-\tfrac{1}{2}}|
                                      + x_\pm y_\pm |n{-3}{-\tfrac{1}{2}}\rangle\langle n2{-\tfrac{1}{2}}| \nonumber \\
                                      && + x_\pm y_\pm |n2{-\tfrac{1}{2}}\rangle\langle n{-3}{-\tfrac{1}{2}}|
\end{eqnarray}
The expansion coefficients $x_\pm$ and $y_\pm$ are obtained from experimental measurements, see main text.
The remaining density matrix operators can be obtained with the help of the time-reversal operator, i.e.
$\hat{\rho}_{\Gamma_7^b} = \hat{\cal T}\hat{\rho}_{\Gamma_7^a}\hat{\cal T}$,
$\hat{\rho}_{\Gamma_{8\pm}^b} = \hat{\cal T}\hat{\rho}_{\Gamma_{8\pm}^a}\hat{\cal T}$,
and $\hat{\rho}_{\Gamma_{8\pm}^d} = \hat{\cal T}\hat{\rho}_{\Gamma_{8\pm}^c}\hat{\cal T}$.
\end{widetext}

\subsection{Atomic d-spinors}
The atomic Dirac spinors read \cite{GreinerBook}
\begin{eqnarray}
        \Psi_{n\kappa m_j}(r\theta\phi) = \frac{1}{r}\begin{bmatrix}
                F_{n\kappa}(r)\Omega_{\kappa m_j}(\theta\phi) \\
                {\rm i}G_{n\kappa}(r)\Omega_{-\kappa m_j}(\theta\phi)
        \end{bmatrix}
\end{eqnarray}
where $r$, $\theta$ and $\phi$  are the spherical polar coordinates, and $F_{n\kappa}(r)$ and $G_{n\kappa}(r)$ are the
large and small radial components, respectively.
The sphreical spinors $\Omega_{\kappa m_j}(\theta\phi) = \langle\theta\phi|km_j\rangle$ are defined in terms
of the spherical harmonics $Y_{\ell m_\ell}(\theta\phi)=\langle\theta\phi|\ell m_\ell\rangle$ by
\begin{equation}
        |\kappa m_j\rangle =
        \begin{bmatrix}
                {\rm sgn}(-\kappa)\sqrt{\frac{\kappa+\frac{1}{2}-m_j}{2\kappa+1}}|\ell,m_j-\frac{1}{2}\rangle\\
                             \sqrt{\frac{\kappa+\frac{1}{2}+m_j}{2\kappa+1}}|\ell,m_j+\frac{1}{2}\rangle
        \end{bmatrix}
\end{equation}
Neglecting the small component $G_{n\kappa}(r)\approx0$ within the approximation $F_{nd_{5/2}}(r)\approx F_{nd_{3/2}}(r)$,
and using the Pauli eigenvectors
\begin{eqnarray}
                |\alpha\rangle = \begin{bmatrix} 1 \\ 0 \end{bmatrix}, \quad
                |\beta\rangle  = \begin{bmatrix} 0 \\ 1 \end{bmatrix}
\end{eqnarray}
        the atomic $nd_{3/2}$-spinors and $nd_{5/2}$-spinors can be recast as
\begin{eqnarray}
        && |n2{-\tfrac{3}{2}}\rangle = \sqrt{\tfrac{1}{5}}|nd_{-1}\beta\rangle - \sqrt{\tfrac{4}{5}}|nd_{-2}\alpha\rangle \\
        && |n2{-\tfrac{1}{2}}\rangle = \sqrt{\tfrac{2}{5}}|nd_0\beta\rangle - \sqrt{\tfrac{3}{5}}|nd_{-1}\alpha\rangle \\
        && |n2{+\tfrac{1}{2}}\rangle = \sqrt{\tfrac{3}{5}}|nd_{+1}\beta\rangle - \sqrt{\tfrac{2}{5}}|nd_0\alpha\rangle \\
        && |n2{+\tfrac{3}{2}}\rangle = \sqrt{\tfrac{4}{5}}|nd_{+2}\beta\rangle - \sqrt{\tfrac{1}{5}}|nd_{+1}\alpha\rangle
\end{eqnarray}
        and
\begin{eqnarray}
        && |n{-3}{-\tfrac{5}{2}}\rangle = |nd_{-2}\beta\rangle \\
        && |n{-3}{-\tfrac{3}{2}}\rangle = \sqrt{\tfrac{4}{5}}|nd_{-1}\beta\rangle + \sqrt{\tfrac{1}{5}}|nd_{-2}\alpha\rangle \\
        && |n{-3}{-\tfrac{1}{2}}\rangle = \sqrt{\tfrac{3}{5}}|nd_{0}\beta\rangle + \sqrt{\tfrac{2}{5}}|nd_{-1}\alpha\rangle \\
        && |n{-3}{+\tfrac{1}{2}}\rangle = \sqrt{\tfrac{2}{5}}|nd_{+1}\beta\rangle + \sqrt{\tfrac{3}{5}}|nd_{0}\alpha\rangle \\
        && |n{-3}{+\tfrac{3}{2}}\rangle = \sqrt{\tfrac{1}{5}}|nd_{+2}\beta\rangle + \sqrt{\tfrac{4}{5}}|nd_{+1}\alpha\rangle \\
        && |n{-3}{+\tfrac{5}{2}}\rangle = |nd_{+2}\alpha\rangle
\end{eqnarray}
respectively. Inserting the above form of the atomic $nd_j$-spinors into the matrix density operators of the
relativistic crystal field states and tracing out the orbital angular momentum degree of freedom, the matrix representation
of the reduced density matrices for the spin halfspace and thereof the eigenvalues required for the von Neumann entropy,
are straightforwardly obtained.

\subsection{Calculation of crystal field parameters}
First, the parameter $\delta$ is obtained by fitting the equation
\begin{widetext}
\begin{eqnarray}
        \left(\mu_{\rm eff}/\mu_{\rm B}\right)^2 &=&
        \frac{3}{200} \left(5 + \frac{(\delta-\sqrt{\delta^2+1})^2 - 1 + 4\sqrt{6}(\delta-\sqrt{\delta^2+1})}{(\delta-\sqrt{\delta^2+1})^2+1}
        \right)^2 \nonumber \\
        &+& \frac{1}{150} \left(\frac{18-33(\delta-\sqrt{\delta+1})^2-2\sqrt{6}(\delta-\sqrt{\delta^2+1})}{(\delta-\sqrt{\delta^2+1})^2+1}
        \right)^2
\end{eqnarray}
\end{widetext}
to the known effective magnetic dipole moment $\mu_{\rm eff}$.
Then, the parameters derived from the relativistic crystal field theory are calculated using the following equations:
\begin{eqnarray}
        && Dq = \left(1 + \frac{\delta}{\sqrt{\delta^2+1}}\right)\frac{\Delta_1}{12}-\frac{\Delta_2}{6} \\
        && \xi_{nd} = \frac{2\delta\Delta_1}{5\sqrt{\delta^2+1}} - \frac{4}{5}Dq \\
        && p/q = \frac{1}{2\delta\sqrt{6}}\left(\frac{5\xi_{nd}}{4Dq} + 1 \right)
\end{eqnarray}
The obtained crystal field strength $Dq$, spin-orbit coupling constant $\xi_{nd}$, and relativistic ratio $p/q$,
reproduce the experimental $d$-$d$ optical transitions energies $\Delta_1$ and $\Delta_2$, and effective magnetic dipole moment
$\mu_{\rm eff}$, according to the relativistic crystal field theory.

\section*{Acknowledgments}
There is no funding to report in this work.
J. F. P\'erez-Torres conceived the original idea, developed the theoretical framework, and performed the calculations.
V. Garc\'ia-Rojas located published experimental data and interpreted the results.
All authors discussed the results and contributed to the final version of the manuscript.

\section*{Data availabity}
All the numerical results shown in this work can be obtained straightfordwardly from the formulas presented
in the main text following the protocol described in the Appendix.

\bibliography{rcft}

\end{document}